\documentclass{appolb}
\usepackage{epsfig}
\usepackage{rotating}
\begin{document}
\title{The Nature of the  $X(2175)$
\thanks{Talk at Workshop \em ``Excited QCD'', \em Zakopane, Poland,
8--14 Feb.\ 2009}
}
\author{S. Coito$^*$, G. Rupp
\address{
Centro de F\'isica das Interac\c c\~oes Fundamentais, Instituto Superior
T\'ecnico, Technical University of Lisbon, P-1049-001 Lisboa, Portugal}
\and
E. van Beveren
\address{
Centro de F\'{\i}sica Computacional, Departamento de F\'{\i}sica,
Universidade de Coimbra, P-3004-516 Coimbra, Portugal}
}

\maketitle

\begin{abstract}
We study the puzzling vector meson X(2175) in a multichannel generalisation
of the Resonance-Spectrum-Expansion model. Besides the usual $P$-wave
pseudoscalar--pseudoscalar, pseudoscalar--vector, and vector--vector channels
that couple to mesons with vector quantum numbers, we also include the
important $S$-wave vector--scalar, pseudoscalar--axial-vector and
vector--axial-vector channels, including the observed $\phi(1020)\,f_0(980)$
decay mode. The strong coupling to nearby $S$-wave channels originates
dynamically generated poles, two of which come out close to the energy region
of the $X(2175)$, viz.\ at $(2.037-i0.170)$ GeV and $(2.382-i0.20)$ GeV.
Further improvements are proposed.
\end{abstract}

\section{Introduction}
The $X(2175)$ was first observed by BABAR \cite{PRD74p091103} in the
process $e^+e^-\rightarrow \phi(1020)f_0(980)$, and identified as a $1^{--}$
resonance, with $M=(2.175\pm0.010\pm0.015)$ GeV and $\Gamma=(58\pm16\pm20)$
MeV. This state was then confirmed by BES and denoted $Y(2175)$
\cite{PRL100p102003}, from the decay $J/\Psi\rightarrow\eta\phi f_0(980)$,
with $M=(2.186\pm0.010\pm0.006)$ GeV and $\Gamma=(65\pm25\pm17)$ MeV. It is
now included in the PDG Particle Listings \cite{PLB667p1} as the $\phi(2170)$.

On the theoretical side, the $X(2175)$ has been descibed as a three-meson
resonance in a Faddeev calculation for the $\phi K \bar{K}$ system
\cite{PRD78p074031}, obtaining a narrow peak around  2150~MeV, about $27$ MeV
wide. Earlier, a conventional Resonance-Chiral-Perturbation-Theory calculation 
\cite{PRD76p074012} failed to produce such a peak, which led to the former
3-body model. Other approaches include QCD-sum-rule calculations for a
tetraquark state \cite{NPA791p106}, and a perturbative multichannel analysis
to distinguish between a strangeonium hybrid and a normal $2\,{}^{3\!}D_1$
$s\bar{s}$ state \cite{PLB657p49}.

\section{Resonance Spectrum Expansion and the $X(2175)$}
In the present study of the $X(2175)$, we use the
Resonance-Spectrum-Expansion (RSE) model \cite{IJTPGTNO11p179} to unitarise
a normal $s\bar{s}$ spectrum. In the RSE approach, non-exotic mesons
are described as regular quark-antiquark states, but non-perturbatively dressed
with meson-meson components. An important feature is the inclusion of a
complete $q\bar{q}$ confinement spectrum in the intermediate state
\cite{IJTPGTNO11p179,0809.1149}, resulting for the multichannel case in an
effective meson-meson potential
\begin{equation}
V_{ij}^{(L_i,L_j)}(p_i,p'_j;E)=\lambda^2\,j^i_{L_i}(p_ia)\,j^j_{L_j}(p'_ja)
{\displaystyle \sum_{n=0}^{\infty} \frac{g_i(n)g_j(n)}{E-E_n} } \; ,
\label{veff}
\end{equation}
where $\lambda$ is an overall coupling constant, $a$ a parameter mimicking
the average string-breaking distance, $j^i_{L_i}$ a spherical Bessel
function, $p_i$ and $p'_j$ the relativistically defined relative momenta
of initial channel $i$ and final channel $j$, respectively, $g_i(n)$ the
coupling of channel $i$ to the $n$-th $q\bar{q}$ recurrence, and $E_n$ the
discrete energy of the latter confinement state. Note that the couplings
$g_i(n)$, evaluated on a harmonic-oscillator (HO) basis for the $^{3\!}P_0$
model \cite{ZPC21p291}, decrease very rapidly for increasing $n$,
so that practical convergence is achieved with the first 20 terms in the
infinite sum. Moreover, the separable form of the effective potential
(\ref{veff}) allows the solution of the (relativistic) Lippmann-Schwinger
equation in closed form.

In the present first study, we restrict ourselves to the $^{3\!}S_1$ channel
for the $q\bar{q}$ system. Moreover, we assume ideal mixing, so that only
$s\bar{s}$ states are considered. For the confinement mechanism, we take
an HO potential. This choice is not strictly necessary, as Eq.~(\ref{veff})
allows for any confinement spectrum, but the HO has shown to work fine in
practically all phenomenological applications. The flavour-dependent HO
spectrum reads
\begin{equation}
E_n=m_q+m_{\bar{q}}+\omega(2n+3/2+\ell) \;.
\label{ho}
\end{equation}
The parameter values $\omega=190$ MeV and $m_s=508$ MeV are kept unchanged
with respect to all previous work (see e.g.\ Ref.~\cite{PRD27p1527}).
In Table~\ref{hos}, we list some of the eigenvalues given by Eq.~(\ref{ho}).
As for the decay sector that couples to $J^{PC}\!=\!1^{--}$ $\phi$ states,
we take all pseudoscalar--pseudoscalar (PP), pseudoscalar--vector (PV),
and vector--vector (VV) channels, which are in $P$-waves,
as well as all vector--scalar (VS), pseudoscalar--axial-vector (PA), and 
vector--axial-vector (VA) channels, being in $S$-waves. These 15 channels are
listed in Table~\ref{MM}, including the observed
\cite{PRD74p091103,PRL100p102003} $\phi f_0(980)$ mode,
with the respective orbital angular momenta, spins, and thresholds.
\begin{table}[t]
\centering
\begin{tabular}{|c c|}
\hline
$n$ & $s\bar{s}$\\ 
\hline
0 & 1.301\\
1 & 1.681\\
2 & 2.061\\
3 & 2.441\\
4 & 2.821 \\
\hline
\end{tabular}
\mbox{} \\[1mm]
\caption{HO eigenvalues (\ref{ho}) in GeV, for $\omega=190$ MeV,
$m_s=508$ MeV, $\ell=0$.}
\label{hos}
\end{table}
\begin{table}[h]
\centering
\begin{tabular}{|l|c|c|}
\hline
Channel & Relative $L$, Total $S$ & Threshold \\ 
\hline
$KK$ & $1,0$ & $0.987$\\
$KK^*$ & $1,1$ &$1.388$\\ 
$\eta\phi$& $1,1$ & $1.567$\\ 
$\eta'\phi$& $1,1$ & $1.977$\\ 
$K^*K^*$&  $1,0$  &$1.788$\\
$K^*K^*$&  $1,2$  &$1.788$\\ 
$\phi f_0(980)$& $0,1$ & $1.999$\\ 
$K^*K_0^*(800)$& $0,1$ &$1.639$\\
$\eta h_1(1380)$ & $0,1$ & $1.928$\\
$\eta' h_1(1380)$& $0,1$  &  $2.338$\\
$KK_1(1270)$& $0,1$  & $1.764$\\
$KK_1(1400)$&  $0,1$ & $1.894$\\
$K^*K_1(1270)$& $0,1$  & $2.164$\\
$K^*K_1(1400)$&  $0,1$  & $2.294$\\
$\phi f_1(1420)$& $0,1$ & $2.439$\\
\hline
\end{tabular}
\mbox{} \\[2mm]
\caption{Thresholds in GeV of included meson-meson channels
(see Ref.~\cite{PLB667p1}).}
\label{MM}
\end{table}

\section{Results}
As emphasised above, the $T$-matrix for the effective potential~(\ref{veff})
can be solved in closed form. Bound states and resonances correspond to poles
of $T$ on the appropriate sheet of the many-sheeted Riemann surface. For the 15
channels considered, there are $2^{15}=32,768$ such sheets. However, the
relevant poles are in principle those that correspond to relative momenta with
negative imaginary parts with respect to open channels, and positive imaginary
parts for closed channels. The simplest example of a latter-type pole is
a bound state, for which the real part of the momentum is zero.

\begin{figure}[t]
\begin{tabular}{lr}
\resizebox{!}{161pt}{\includegraphics{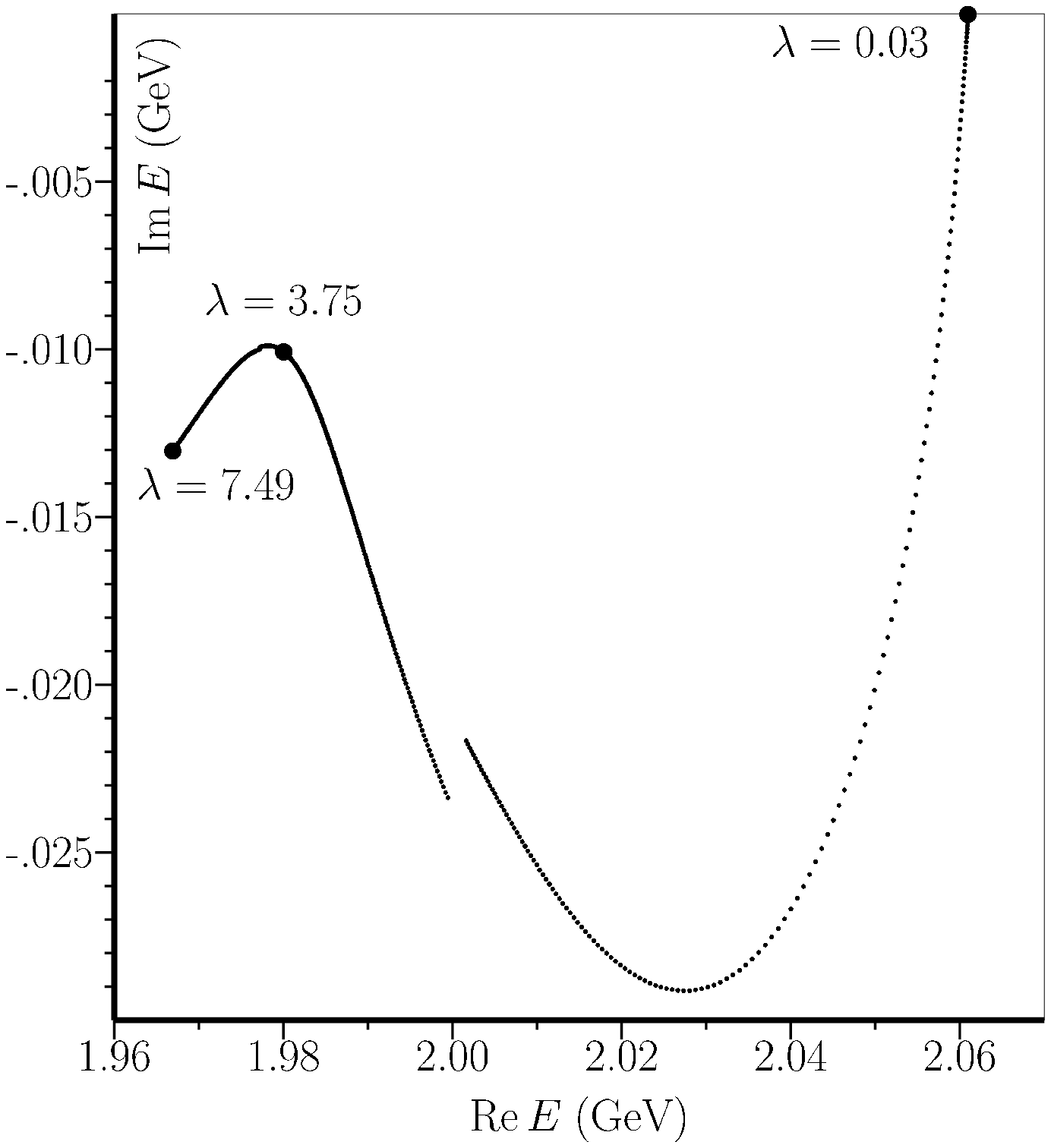}}
&
\hspace*{25pt}\resizebox{!}{161pt}{\includegraphics{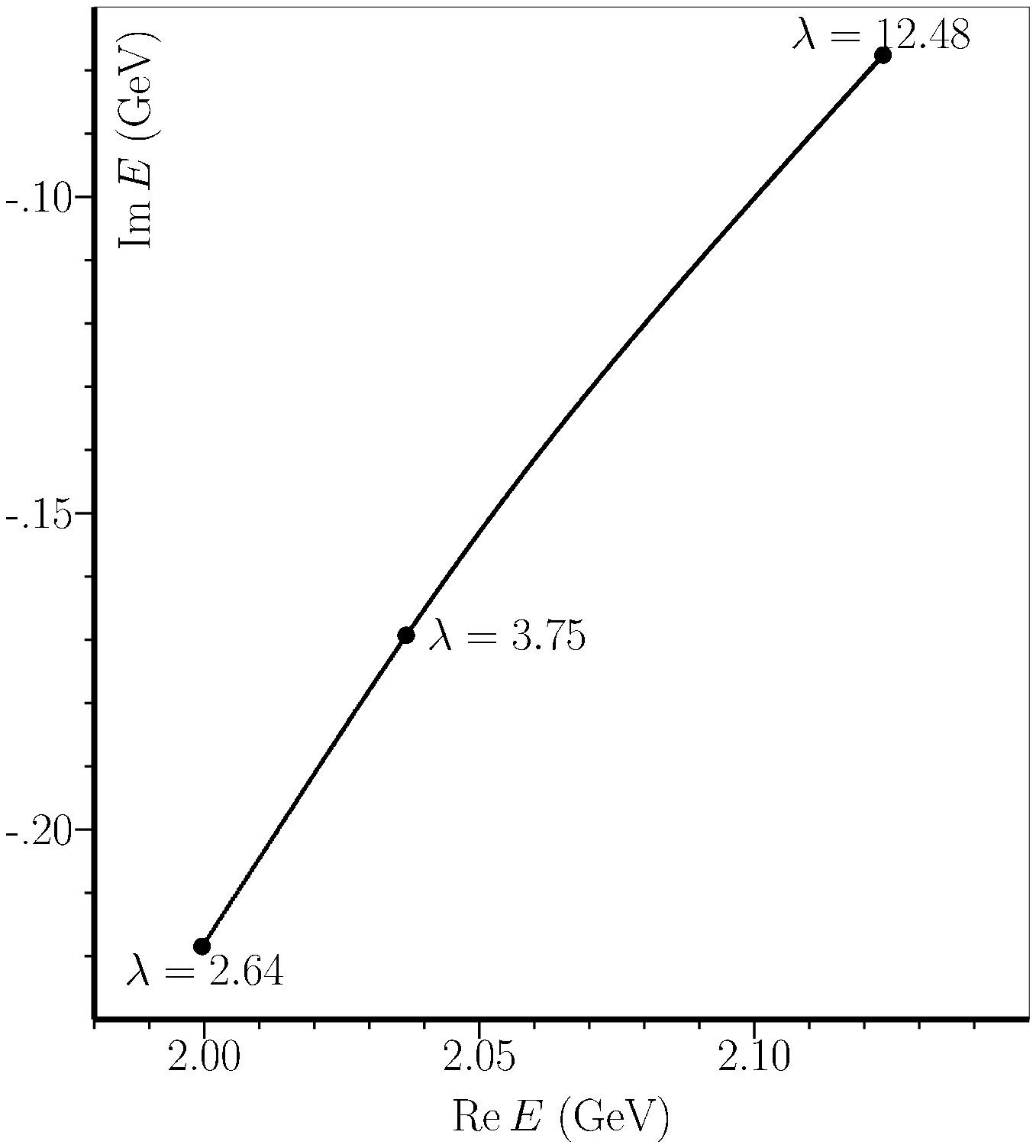}}
\end{tabular}
\caption{Confinement pole for $3\,{}^{3\!}S_1$ state (left); first
continuum pole (right).}
\label{traj1}
\end{figure}

\begin{figure}[h]
\begin{tabular}{lr}
\resizebox{!}{161pt}{\includegraphics{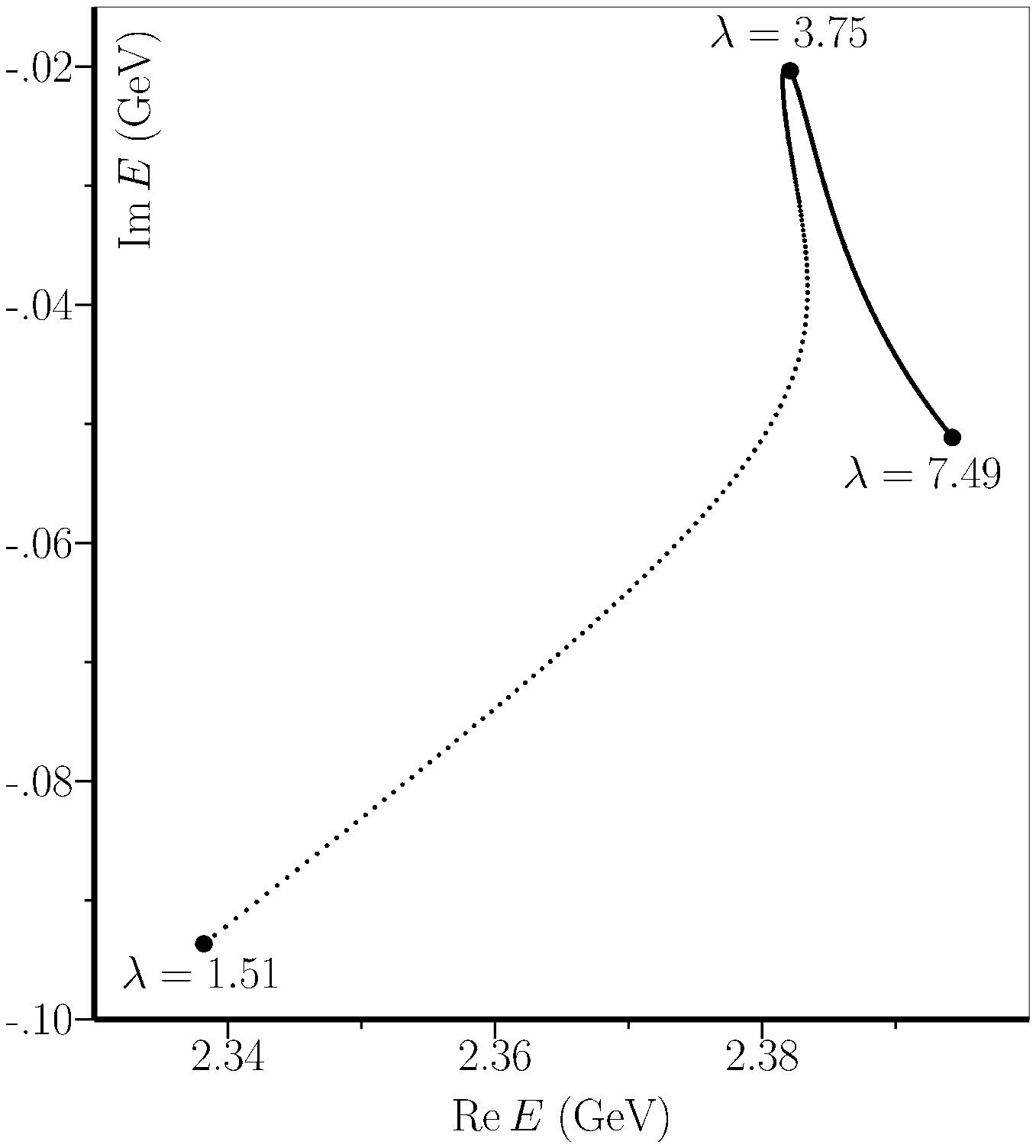}}
&
\hspace*{25pt}\resizebox{!}{161pt}{\includegraphics{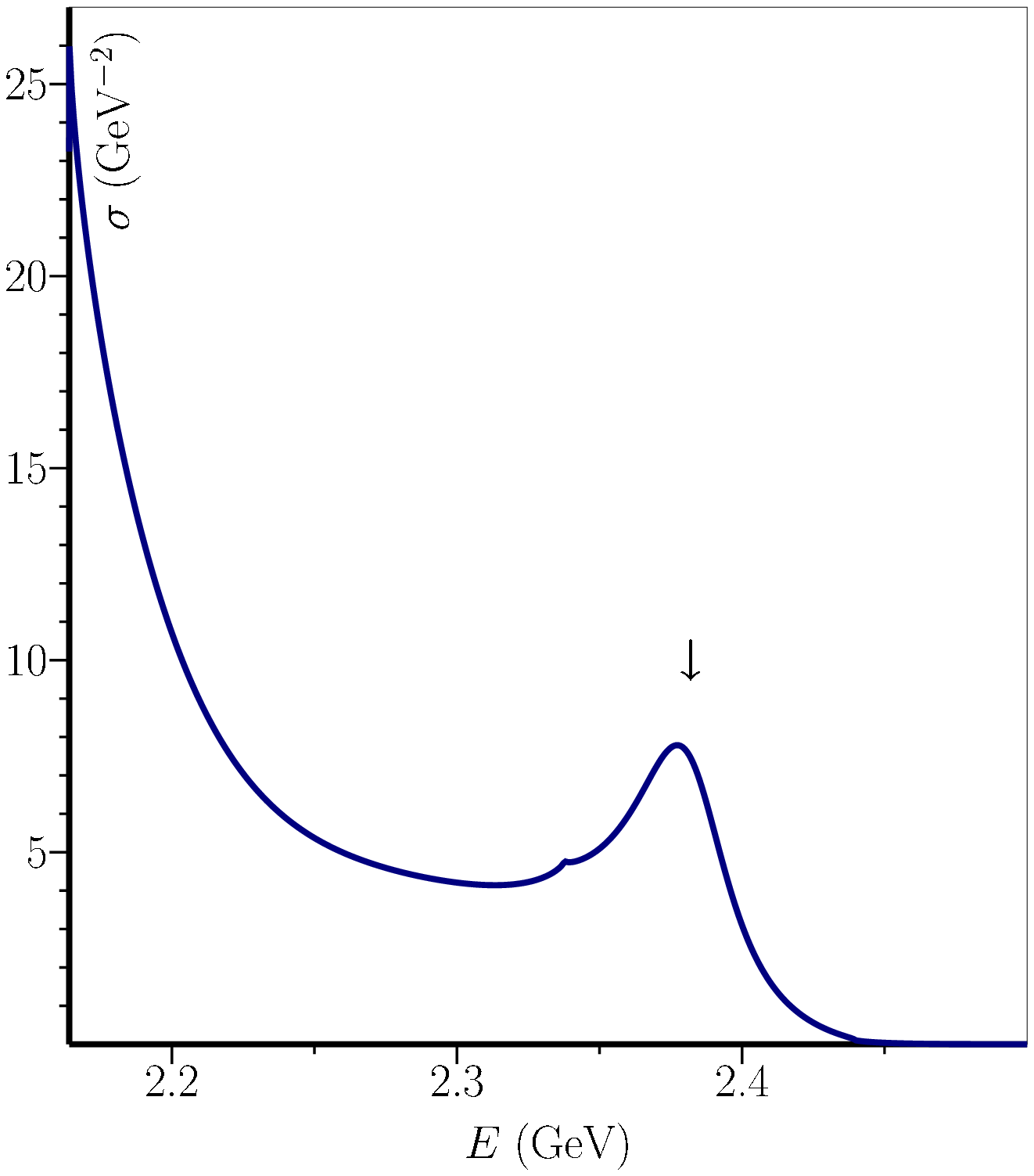}}
\end{tabular}
\caption{Second continuum pole (left); elastic $K^*K_1(1270)$ cross
section (right).}
\label{traj2cross3}
\end{figure}

The only two free parameters of the model, viz.\ $\lambda$ and $a$, we fix by
demanding that the mass and width of the $\phi(1020)$ be reasonably reproduced.
For {\boldmath$a=5.0$} GeV$^{-1}$ and {\boldmath$\lambda=3.75$} GeV$^{-3/2}$ we
get a theoretical pole position of
{\boldmath$E_{\mbox{\scriptsize\bf theor}}=(1.0145 - i 0.0034)$}~GeV, to be
compared with the PDG \cite{PLB667p1} value
{\boldmath$E_{\mbox{\scriptsize\bf exp}}=(1.0195-i0.0021)$}~GeV, which is more
than good enough for the present simplified investigation. Of course, besides
the $\phi(1020)$, there are several other resonance poles, which are given
in Table~\ref{poles}, for energies up to about 2.5~GeV\footnote{These results
are slightly different from the preliminary ones presented at the workshop,
due to a, now corrected, minor error in the computer code.}.
\begin{table}[ht]
\centering
\begin{tabular}{|c|c|c|}
\hline
$\Re$e & $\Im$m & Type of Pole \\ 
\hline
$1.0145$&$-0.0034$& confinement, $n=0$\\
$1.457$ & $-0.011$& confinement, $n=1$\\	
$1.980$ & $-0.010$& confinement, $n=2$\\	
$2.037$ & $-0.170$& continuum\\		
$2.382$ & $-0.020$& continuum\\
$2.580$ & $-0.123$& continuum\\		 
\hline
\end{tabular}
\mbox{} \\[1mm]
\caption{Pole positions, in GeV.}
\label{poles}
\end{table}
Confinement poles are those that end up at the energies of the confinement
spectrum in the limit $\lambda\to0$. These usually have small to moderate
imaginary parts. On the other hand, the continuum poles, which are dynamically
generated, disappear in the complex energy plane for $\lambda\to0$, with
$\Im\mbox{m}E\to-\infty$. The latter poles mostly have large imaginary parts,
for physical values of $\lambda$, but there are exceptions, like the fifth
pole in Table~\ref{poles}. Typical cases of pole trajectories are shown in
Fig.~\ref{traj1}, with the $n\!=\!2$ confinement pole in the left-hand plot and
the first continuum pole in the right-hand one. The jump in the trajectory of
the confinement pole is due to a change of Riemann sheet at the $\phi f_0(980)$
threshold. The trajectory of the second continuum pole is depicted in
Fig.~\ref{traj2cross3}, left-hand plot, which shows its highly non-linear and
non-perturbative behaviour. As for the main purpose of the present work, we
find two poles in the energy region 2.0--2.4~GeV relevant for the $X(2175)$,
namely at {\boldmath$(2.037-i0.170)$}~GeV and {\boldmath$(2.382-i0.020)$}~GeV,
being both continuum poles. The $4\,{}^{3\!}S_1$ ($n\!=\!3$)
confinement pole can be easily followed from $E=2.441$~GeV, up to a value of
$\lambda\approx3.1$~GeV$^{-3/2}$, but one loses its track when switching
Riemann sheet at the opening of the $\phi f_1(1420)$ channel. In any case, all
these precise pole positions are not so important for this first study. Suffice
it to say that dynamical poles can be generated in the energy region pertinent
to the $X(2175)$, and possibly with quite small imaginary parts. Preliminary
results for a more complete calculation \cite{preparation}, including the
$^{3\!}D_1$ states, indicate a considerable improvement of the pole positions
for the $X(2175)$ candidates.

Having the exact $T$-matrix at our disposal, we can easily compute
observables, too, like cross sections and phase shifts. Although this is
not of great importance here, in view of the rather tentative pole positions
above 2~GeV, we show in Fig.~\ref{traj2cross3}, right-hand plot, the $S$-wave
$K^*K_1(1270)$ cross section, just as an illustration. The pole at
$(2.382-i0.020)$~GeV is clearly visible here, in contrast with the
$\phi f_0(980)$ channel. Inclusion of the $^{3\!}D_1$ states turns out to
significantly improve the description, both for the pole positions and for the
$\phi f_0(980)$ cross section \cite{preparation}, in line with experiment.
 
\section{Conclusions and outlook}
We have shown that coupling a spectrum of confined $s\bar{s}$ states to all
$S$-wave and $P$-wave two-meson channels composed of light mesons allows to
generate dynamical resonances above 2~GeV, besides roughly reproducing the
mass and the width of the $\phi(1020)$. This may provide a framework to
understand the puzzling $X(2175)$ meson, owing to the large and
non-linear coupled-channel effects, especially from the $S$-wave channels.
Inclusion of the $^{3\!}D_1$ $s\bar{s}$ states will then account for a more
realistic modelling, as confirmed by preliminary results \cite{preparation}.
Further improvements may be considered as well, such as deviations from ideal
mixing, smearing out resonances in the final state, and more general
transition potentials.

\section*{Acknowledgements}
We thank the organisers for an inspiring and pleasant workshop. We are
also indebted to K.~Khemchandani for very useful discussions.
This work was supported in part by
the \emph{Funda\c{c}\~{a}o para a Ci\^{e}ncia e a Tecnologia}
\/of the \emph{Minist\'{e}rio da Ci\^{e}ncia, Tecnologia e Ensino Superior}
\/of Portugal, under contract CERN/\-FP/\-83502/\-2008.

\end{document}